\documentclass[aps,prb,amsmath,amssymb,twocolumn,floatfix]{revtex4-1}
\usepackage{graphicx}% Include figure files
\usepackage{dcolumn}% Align table columns on decimal point
\usepackage{bm}% bold math

\begin{document}

\title{Quantum Spin Fluctuations in the Bulk Insulating State of Pure and Fe-doped SmB$_6$}

\author{K.~Akintola,$^1$ A.~Pal,$^1$ M.~Potma,$^{1,2}$ S.R.~Saha,$^3$ X.F.~Wang,$^3$ J.~Paglione,$^{3,4}$ and J.E.~Sonier,$^{1,4}$}

\affiliation{$^1$Department of Physics, Simon Fraser University, Burnaby, British Columbia V5A 1S6, Canada \\
$^2$Kwantlen Polytechnic University, Richmond, British Columbia V6X 3X7, Canada \\
$^3$Center for Nanophysics and Advanced Materials, Department of Physics, University of Maryland, College Park, Maryland 20742, USA \\
$^4$Canadian Institute for Advanced Research, Toronto, Ontario M5G 1Z8, Canada}

\date{\today}
%%%%%%%%%%%%%%%%%%%%%%%%%%%%%%%%%%%%%%%%%%%%%%%%%%%%%%%
\begin{abstract}
The intermediate-valence compound SmB$_6$ is a well-known Kondo insulator, in which hybridization of itinerant $5d$ electrons with localized $4f$ electrons 
leads to a transition from metallic to insulating behavior at low temperatures. Recent studies suggest that SmB$_6$ is a topological insulator, with topological 
metallic surface states emerging from a fully insulating hybridized bulk band structure. Here we locally probe the bulk magnetic properties of pure and 
0.5~\% Fe-doped SmB$_6$ by muon spin rotation/relaxation ($\mu$SR) methods. Below 6 K the Fe impurity induces simultaneous changes in the bulk local 
magnetism and the electrical conductivity. In the low-$T$ insulating bulk state we observe a temperature-independent dynamic relaxation rate indicative 
of low-lying magnetic excitations driven primarily by quantum fluctuations.
\end{abstract}

\maketitle
%%%%%%%%%%%%%%%%%%%%%%%%%%%%%%%%%%%%%%%%%%%%%%%
Topological insulators are exotic quantum states of matter characterized by an electrically insulating bulk and topologically-protected metallic surface states. 
Due to an interplay of strong correlations and strong spin-orbit coupling of the $4f$ electrons, SmB$_6$ is predicted to develop a non-trivial $\mathbb{Z}_2$ 
topological insulating state.\cite{Dzero:16} Angle-resolved photoemission\cite{Xu:14} and point-contact spectroscopy\cite{Zhang:13} measurements show that 
the crossover from the bulk high-$T$ metallic state to the low-$T$ Kondo insulating phase occurs gradually over a fairly wide temperature range 
(30~K~$\! < T \! < \! 110$~K). Transport measurements show that surface electrical conduction occurs below $T \! \sim \! 5$ to 6~K with a resistance that 
saturates at lower temperature.\cite{Kim:13,Wolgast:13,Kim:14} The low-$T$ conduction arises from two-dimensional states\cite{Li:14} that occur in the hybridization 
gap exclusively at the surface,\cite{Zhang:13,Kim:13,Syers:15} as expected for metallic surface states of topological origin.\cite{Nakajima:16} Yet the ground 
state of SmB$_6$ is still unclear, in part because not all bulk properties at low $T$ are that of a conventional band-gapped insulator. Despite the loss of 
bulk electrical conduction, quantum oscillations consistent with a bulk Fermi surface have been observed,\cite{Tan:15} and the low-temperature specific heat exhibits 
a significant bulk residual $T$-linear term typical of a metallic state.\cite{Wakeham:16} Recently, it has been argued that there is some residual bulk electrical 
conductivity in SmB$_6$ below 4~K.\cite{Gabani:15} There also exists significant bulk ac-conduction arising from low-energy states in the Kondo gap.\cite{Laurita:16} 

Nuclear magnetic resonance (NMR) Knight shift and spin-lattice relaxation rate ($1/T_1$) measurements,\cite{Caldwell:07} bulk magnetic susceptibility,\cite{Glushkov:06} 
Raman spectroscopy,\cite{Nyhus:97,Valentine:16} and inelastic neutron scattering (INS)\cite{Alekseev:95,Fuhrman:15} studies of SmB$_6$ reveal the emergence of 
bulk in-gap bound states of a different origin below $T \! \sim \! 20$-30~K. The sharp dispersive magnetic excitations observed at 14 meV within the hybridization 
gap by INS have been attributed to a bulk collective spin exciton resonance mode due to residual antiferromagnetic (AFM) quasiparticle 
interactions.\cite{Riseborough:00,Riseborough:03} These bound magnetic quasiparticle states are robust due to the protection provided by the hybridization gap, 
and there is evidence that the spin excitons couple to bulk in-gap states introduced by disorder.\cite{Valentine:16} Theoretical \cite{Kapilevich:15} 
and planar tunnelling spectroscopy\cite{Park:16} studies suggest there is an incomplete protection of the surface states of SmB$_6$ due to interactions of these 
bulk spin excitons with the surface states.

Previous $\mu$SR studies of SmB$_6$ detected fluctuating electronic moments in the bulk of floating-zone (FZ)
grown single crystals down to $T \! = \! 0.019$~K, characterized by a zero-field muon spin relaxation rate ($\lambda_{\rm ZF}$) that exhibits a 
small distinct peak near 5~K and saturates below 1~K.\cite{Biswas:14,Biswas:17} A similar, but significantly broader ($\sim \! 3$ times wider) peak is observed in the temperature 
dependence of $1/T_1$ by NMR, which weakens and shifts to higher $T$ with increasing applied magnetic field $H$.\cite{Caldwell:07}
The magnetic fluctuations observed in these studies have been attributed to bulk magnetic in-gap states, although the pronounced low-temperature 
peak in $\lambda_{\rm ZF}$ has not been explained. 

Here we report low-temperature (0.024~K~$\leq \! T \! \leq \! 16$~K) $\mu$SR measurements on pure and 0.5~\% Fe-doped SmB$_6$ single crystals in zero field (ZF), 
longitudinal field (LF), and weak transverse field (wTF) configurations performed at TRIUMF in Vancouver, Canada.
In all cases the initial muon spin polarization {\bf P}(0) was antiparallel to the direction of the linear momentum of the positive muon beam (defined to be
the $z$-direction) and parallel to the crystallographic $c$-axis. 
While magnetic impurities can induce time-reversal symmetry breaking and open up a gap at the Dirac point of a topological insulator,\cite{Chen:10} 
the current study is concerned with the effects of Fe impurities on the bulk magnetic properties and the relationship between the magnetism and 
electrical conductivity. 

The single crystals of pure and 0.5~\% Fe-doped SmB$_6$ were grown by an Al flux method.\cite{Nakajima:16} 
Single crystals of SmB$_6$ grown by the Al-flux method were previously only studied by $\mu$SR down to 2~K and shown to have a substantially 
lower ZF relaxation rate below 15~K compared to FZ grown single crystals.\cite{Biswas:14} This difference is likely
a result of Sm vacancies in the FZ crystals.\cite{Phelan:16}  
Stoichiometric quantities of Sm chunk, Fe powder and B powder as the reactants, and Al as the flux were carefully ground in a 
ratio of (Sm, Fe)B$_6$:Al = 1:200. The starting materials were placed in an alumina crucible in a tube furnace 
and then pumped to high vacuum and purged with Ar three times. With a slow Ar flow through the tube, the mixture was heated up to 1600~$^\circ$C, kept for 2 days, and
then slowly cooled down to 600~$^\circ$C at a rate of -2~$^\circ$C/h. The Al flux was removed by concentrated NaOH etching. High quality and mostly cubic-shaped single 
crystals with typical size of approximately 1~mm~$\times$~1~mm~$\times1$~mm were obtained. Single-crystal X-ray diffraction yielded excellent refinement with 
$R \! = \! 0.41$~\%. Elemental and uniform Fe concentrations were confirmed by wavelength dispersive spectroscopy.

\begin{figure}
\centering
\includegraphics[width=8.0cm]{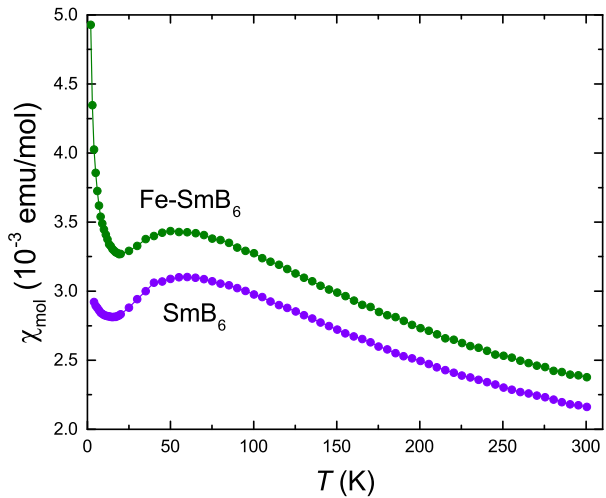}
\caption{(Color online) Temperature dependence of the bulk magnetic susceptibility for a magnetic field of 1~kOe applied parallel 
to the $c$-axis.}
\label{fig1}
\end{figure}

\begin{figure}
\centering
\includegraphics[width=9.5cm]{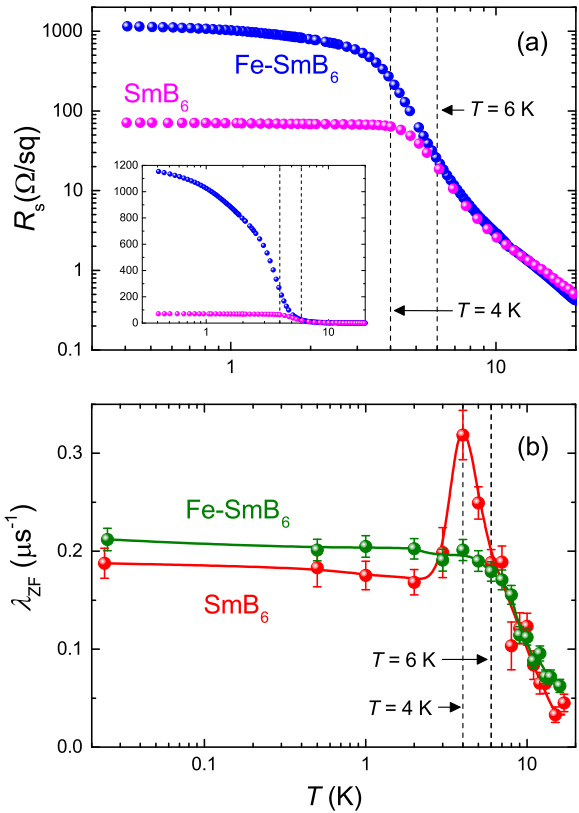}
\caption{(Color online) (a) Temperature dependence of the sheet resistance for both samples at $T \! \geq \! 0.35$~K. The inset shows the same data with a linear vertical scale. 
(b) Temperature dependence of the ZF-$\mu$SR relaxation rate $\lambda_{\rm ZF}$ at $T \! \geq \! 0.024$~K. The solid curves through the data points 
are guides to the eye.}
\label{fig2}
\end{figure}

Figure~\ref{fig1} shows the variation of the bulk magnetic susceptibility $\chi_{\rm mol}$ with temperature for the pure and Fe-doped samples.
The Sm ions fluctuate between non-magnetic Sm$^{2+}$ ($4f^6$) and magnetic Sm$^{3+}$ (4$f^5$) electronic configurations, 
with a Sm$^{3+}$:Sm$^{2+}$ mixed-valence ratio of roughly 6:4 at room temperature.\cite{Mizumaki:09}
Above $T \! \sim \! 110$~K, $\chi_{\rm mol}$ exhibits Curie-Weiss behavior indicative of paramagnetic Sm ions. The broad hump at lower
temperature is a feature of the indirect hybridization gap. Below $T \! \sim \! 15$~K there is a Curie-like upturn in $\chi_{\rm mol}(T)$, which
is generally attributed to paramagnetic impurities.\cite{Gabani:02,Roman:97} In the Fe-doped sample 
the low-$T$ upturn in $\chi_{\rm mol}$ is enhanced and begins at a slightly higher temperature.
Figure~\ref{fig2}(a) shows the temperature dependence of the electrical sheet resistance. In SmB$_6$ the crossover from 
high-$T$ thermally-activated behavior to a low-$T$ plateau has been interpreted as a crossover from an electrically conducting bulk to a state where
electrical conduction occurs predominantly at the surface.\cite{Syers:15} The effect of the Fe impurity on the low-$T$ resistance plateau 
suggests that the Fe suppresses the surface conductance below $T \! \sim \! 6$~K, but does not affect the bulk conductance at higher $T$. 

The ZF-$\mu$SR asymmetry spectrum is defined as $A(t) \! = \! a_0 P_z (t)$, where $a_0$ is the initial asymmetry and $P_z (t)$ is the time evolution 
of the muon-spin polarization in the $z$ direction. The ZF-$\mu$SR spectra for both samples were fit to the sum of sample and background contributions
\begin{equation}
A(t) \! = \! a_0 f G_z(t) + a_0 (1-f) G_{\rm KT}(\Delta_{\rm B} ,t) \, ,
\end{equation}
where $f$ is the fraction of recorded muon-decay events associated with positive muons stopping in the sample,
and $G_{\rm KT}(\Delta_{\rm B} ,t)$ is a $T$-independent static Gaussian Kubo-Toyabe function\cite{Uemura:99} that describes 
the background contribution to the signal relaxation. As in Ref.~\onlinecite{Biswas:14}, the relaxation of 
the ZF-$\mu$SR asymmetry spectrum due to the sample is well described by
\begin{equation}
G_z(t) = G_{\rm KT}(\Delta, t) e^{-(\lambda_{\rm ZF} t)^\beta} \, ,
\label{eqn:AsyZF}
\end{equation}
where the static Gaussian Kubo-Toyabe function $G_{\rm KT}(\Delta, t)$ describes $T$-independent relaxation due to nuclear dipole fields, 
and the stretched-exponential function accounts for relaxation by local fields generated by electronic moments. Global fits over the full temperatures range
with the relaxation rate $\lambda_{\rm ZF}$ as a variable parameter
yield $\Delta \! = \! 0.299(6)$~$\mu$s$^{-1}$, $\beta \! = \! 0.79(2)$ for SmB$_6$, and $\Delta \! = \! 0.288(4)$~$\mu$s$^{-1}$, 
$\beta \! = \! 0.79(2)$ for the Fe-doped sample. These values are comparable to those obtained in Ref.~\onlinecite{Biswas:14}.
As shown in Fig.~\ref{fig2}(b), $\lambda_{\rm ZF}$ first increases with decreasing $T$, which is indicative
of a slowing down of local moment fluctuations. Below 6~K, both the temperature dependence of
$\lambda_{\rm ZF}$ and the sheet resistance in the pure and Fe-doped compounds begin to diverge. In SmB$_6$, $\lambda_{\rm ZF}$ exhibits a distinct peak near 
4~K and saturates below 3~K. No coherent oscillation of the ZF-$\mu$SR signal indicative of magnetic order is observed in either sample down to 0.024~K.

\begin{figure}
\begin{center}
\includegraphics[width=8.0cm]{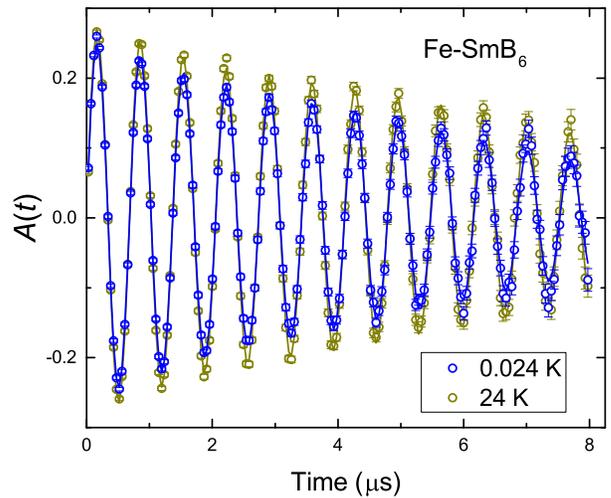}
\end{center}
\caption{Comparison between the wTF-$\mu$SR spectra ($H \! = \! 105$~G) for Fe-doped SmB$_6$ at $T \! =  \! 0.024$~K and 24 K. The solid curves through 
the data points are fits to the sum of two Gaussian-damped cosine functions, corresponding to sample and background contributions.}
\label{fig3}
\end{figure}

Measurements in a wTF oriented perpendicular to the initial muon spin polarization {\bf P}(0) can provide information on the magnetic 
volume fraction. Muons stopping in non-magnetic regions experience a narrow distribution of field due to the nuclear dipoles. Consequently, they contribute 
a weakly-damped component to the wTF-$\mu$SR signal oscillating at a frequency corresponding to the applied magnetic field. Muons 
stopping in magnetic regions experience a broad field distribution associated with the electronic moments, and hence contribute a rapidly-damped component. 
This component can be damped out in the dead time of the spectrometer if the field distribution is sufficiently broad, resulting in a loss of amplitude. 
As shown in Fig.~\ref{fig3}, the wTF-$\mu$SR signal for the Fe-doped SmB$_6$ sample shows no reduction in amplitude or the appearance of a fast relaxing 
component at low temperatures. Consistent with the findings in Ref.~\onlinecite{Biswas:14}, the same is also true for the pure compound. 
Hence we find no evidence of the Fe impurity inducing phase separation into magnetic and non-magnetic regions.

\begin{figure}
\begin{center}
\includegraphics[width=9.5cm]{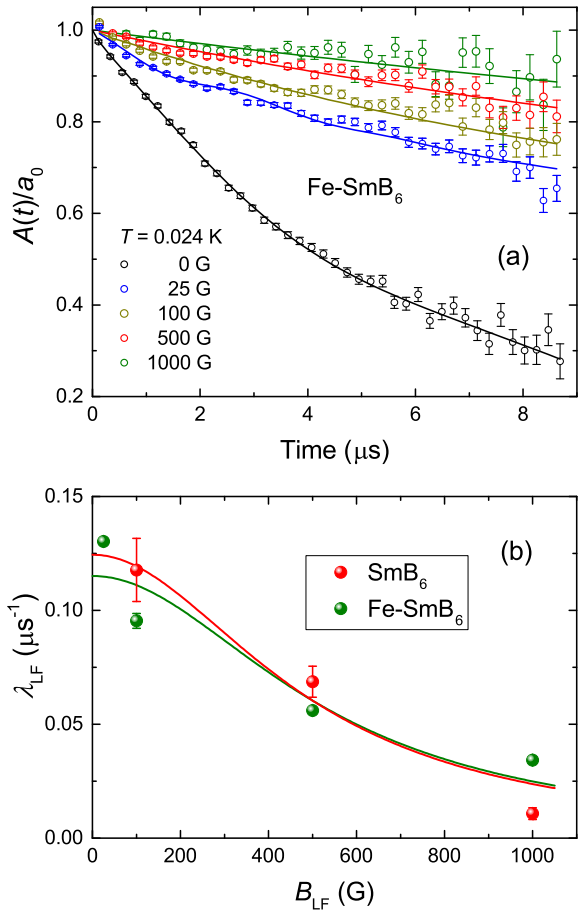}
\end{center}
\caption{(a) Normalized LF-$\mu$SR asymmetry spectra for Fe-doped SmB$_6$ at $T \! =  \! 0.024$~K and for different applied magnetic field strengths.
The curves through the data are fits to Eq.~(\ref{eqn:LF}). 
(b) Dependence of the LF exponential relaxation rate $\lambda_{\rm LF}$ on the magnitude of the applied longitudinal magnetic field. 
The solid (dashed) curve is a fit of the data for 
the Fe-doped (pure) sample to Eq.~(\ref{eqn:Redfield}), which yields $\tau \! = \! 2.2(6) \times 10^{-8}$~s ($2.4(6) \! \times \! 10^{-8}$~s).} 
\label{fig4}
\end{figure}

Longitudinal-field (LF) measurements with the external magnetic field applied parallel to {\bf P}(0) were also performed at 0.024~K to 
determine whether the internal magnetic fields are static or dynamic. Figure~\ref{fig4}(a) shows LF-$\mu$SR asymmetry spectra for the Fe-doped sample
along with fits to the following function
\begin{eqnarray}
A(t) \! & = & \! a_0 [f G_{\rm KT}(B_{\rm LF}, \Delta_{\rm S}, t)e^{-\lambda_{\rm LF} t} \nonumber \\
&  & + (1-f) G_{\rm KT}(B_{\rm LF}, \Delta_{\rm B} ,t)] \, .
\label{eqn:LF}
\end{eqnarray}
Here $G_{\rm KT}(B_{\rm LF}, \Delta_i, t)$ is a Gaussian LF Kubo-Toyabe function,\cite{Uemura:99} which accounts for 
the sample (S) and background (B) nuclear-dipole contributions to the LF signal relaxation. 
Note that fairly good fits are achieved assuming the decay of 
the muon spin polarization by electronic moments in the sample is described by a pure exponential relaxation function, rather than the 
stretched-exponential relaxation assumed in the analysis of the ZF-$\mu$SR spectra. The field dependence 
of the exponential relaxation rate $\lambda_{\rm LF}$ from these fits is shown in Fig.~\ref{fig4}(b). For a Gaussian distribution of local fields typical of a dense 
system of randomly oriented moments, $\lambda_{\rm LF}$ has the following Lorentzian-type dependence on the longitudinal field \cite{Schenck:1985}
\begin{equation}
\lambda_{\rm LF} \! = \! \frac{\lambda_{\rm ZF}}{1 + \left(\gamma_\mu B_{\rm LF} \tau \right)^2}
\! = \! \frac{2 \gamma_\mu^2 \langle B_\mu^2 \rangle \tau}{1 + \left(\gamma_\mu B_{\rm LF} \tau \right)^2} \, ,
\label{eqn:Redfield}
\end{equation}
where $\gamma_\mu/2 \pi$ is the muon gyromagnetic ratio, $\langle B_\mu^2 \rangle$ is the width of the local Gaussian field distribution experienced by 
the muons, and $1/\tau$ is the fluctuation rate of the local field $B_\mu$. The curves in Fig.~\ref{fig4}(b) are fits to Eq.~(\ref{eqn:Redfield}), which
yield a correlation time on the order of $10^{-8}$~s for both samples. Together the LF and wTF measurements at 0.024~K indicate the presence of fluctuating 
electronic moments throughout the entire sample volume. 

The above results are qualitatively similar to the findings in Ref.~\onlinecite{Biswas:14} and establish that a low-$T$ saturation of 
$\lambda_{\rm ZF}$ also occurs in Al-flux grown single crystals. We note that while the low-$T$ saturation in $\lambda_{\rm ZF}$ resembles 
the plateau in the sheet resistance of the pure compound (Fig.~\ref{fig2}), our $\mu$SR experiments are insensitive to the surface and 
probe only the bulk. Persistent spin dynamics are also observed in the small hybridization gap Kondo insulator YbB$_{12}$, 
where weak electronic moments fluctuating at a constant rate have been detected below 4~K by $\mu$SR.\cite{Yaouanc:99}
Like SmB$_6$, the small hybridization gap also plays a role in the occurrence of spin-exciton formation in YbB$_{12}$.\cite{Akbari:09}

The average thermal energy below 30~K is less than 2.6~meV, and below 6~K is less than 0.5~meV. Hence
while previously attributed to intrinsic magnetic in-gaps states,\cite{Biswas:14,Biswas:17} the low-$T$ dynamic spin fluctuations in SmB$_6$ 
appear distinct from the 14 meV collective in-gap mode observed by INS.\cite{Alekseev:95,Fuhrman:15}
Recently, a lower-energy ($\! \lesssim \! 1$~meV) spin exciton branch has been predicted to occur in SmB$_6$ and used to explain
low-$T$ thermodynamic and transport anomalies in SmB$_6$.\cite{Knolle:17}
The persistent spin dynamics observed in pure and Fe-doped SmB$_6$ may be associated with this low-energy spin exciton mode,  
although the saturation of $\lambda_{\rm ZF}$ is uncharacteristic of thermal spin excitations. Instead it
suggests that there are strong quantum effects that are primarily responsible for the low-$T$ spin fluctuations.   

The results in Fig.~\ref{fig2}(b) show that the pronounced peak in $\lambda_{\rm ZF}$ at $T \! \sim \! 4$~K is completely suppressed by the Fe impurity.   
In the pure compound the peak in $\lambda_{\rm ZF}$ begins at 6~K, below which the Fe impurity has an adverse effect on the electrical conductivity.
One possibility is that the critical slowing down of spin fluctuations that is responsible for the increase in $\lambda_{\rm ZF}$ with decreasing $T$,
terminates due to a loss of a conduction-electron mediated RKKY interaction between localized Sm moments when the
bulk conductivity becomes negligible below 4~K. A non-negligible RKKY interaction is suggested by $^{149}$Sm nuclear forward scattering of synchrotron 
radiation and specific heat experiments on SmB$_6$, which detect fluctuating short-range magnetic correlations 
and a pressure-induced magnetically ordered state below 12~K.\cite{Barla:05} In this scenario, the absence of a peak in 
the Fe-doped sample may be due to impurity scattering of conduction electrons. Such scattering may prematurely interrupt an RKKY interaction 
between partially Kondo-compensated Sm moments below 6~K. 

While it is desirable to connect the $\lambda_{\rm ZF}$ peak in SmB$_6$ to the much broader field-dependent $^{11}$B NMR $1/T_1$ peak observed  
at higher temperatures,\cite{Caldwell:07} only the latter is associated with Korringa relaxation. Typically Korringa relaxation is unmeasurably
slow on the $\mu$SR timescale and is ruled out here by the dependence of the muon-spin relaxation rate on LF.
The broad NMR $1/T_1$ peak is suppressed and shifts to higher temperature with increasing $H$.  
Since SmB$_6$ exhibits a negative magnetoresistance at 2~K~$ \! < \! T \! < \! 16$~K indicative of a partial recovery of 
charge carriers from Kondo screening of the Sm moments,\cite{Chen:15} it is clear that the peak in $1/T_1$ cannot be explained by RKKY interactions 
between localized $4f$ moments. In Ref.~\onlinecite{Caldwell:07} it was shown that the field-dependent $1/T_1$ maximum can be explained by
in-gap magnetic states that shift closer to the bottom of the conduction band with increasing $H$. 
Since the Kondo gap is reduced by field,\cite{Jaime:00} there is a diminished protection of the spin excitons with 
increasing $H$ and this may explain why the $1/T_1$ maximum in SmB$_6$ broadens 
and is ultimately wiped out by the field. 

It is possible that the decrease in $\lambda_{\rm ZF}$ below the low-$T$ maximum results from a substantial weakening of exchange coupling 
between conduction electrons and the spin excitons. However, it remains an open question as to why Fe doping causes the peak in the temperature 
dependence of $\lambda_{\rm ZF}$ to vanish. Future INS and theoretical studies are needed to determine the effect of Fe impurities 
on the spin excitons.
  
In summary, the observed saturation of $\lambda_{\rm ZF}$ at low $T$ suggests there are persistent spin dynamics primarily caused by quantum rather 
than thermal fluctuations. This is consistent with a ground state that is in close proximity 
to an AFM quantum critical point. The energy scale of the dynamic spin fluctuations reported here is much lower than the 14~meV coherent resonant mode 
that has been observed by INS, but may relate to a recently predicted low-energy spin exciton branch in SmB$_6$.  

\begin{acknowledgments}
We thank J. Maciejko, P. Riseborough and S. Dunsiger for informative and insightful discussions. J.E.S. acknowledges support from the Natural Sciences and 
Engineering Research Council of Canada. Research at the University of Maryland was supported by AFOSR  
through Grant No. FA9550-14-1-0332 and the Gordon and Betty Moore Foundation's EPiQS Initiative through Grant No. GBMF4419.
\end{acknowledgments}

\end{document}